\pdfoutput=1 
\documentclass{JINST}
\usepackage{upgreek}
\usepackage{subfigure}

\title{Test Beam Campaigns for the CMS Phase I Upgrade Pixel Readout Chip}

\author{S. Spannagel\footnote{for the CMS Collaboration}\\
  DESY Hamburg,\\
  Notkestra\ss e 85, 22607 Hamburg, Germany\\
E-mail: \email{simon.spannagel@desy.de}}

\abstract{The current CMS silicon pixel detector as the innermost component of the CMS experiment is performing well at LHC design luminosity, but would be subject to severe inefficiencies at LHC peak luminosities of $2\times10^{34}\,\mathrm{cm^{-2}s^{-1}}$.

Therefore, an upgrade of the CMS pixel detector is planned, including a new readout chip. The chip design comprises additional on-chip buffer cells as well as high-speed data links and low-threshold comparators in the pixel cells. With these changes the upgraded pixel detector will be able to maintain or even improve the efficiency of the current detector at the increased requirements imposed by high luminosities and pile-up.

The effects of these design changes on e.g. position resolution and charge collection efficiency were studied in detail using a precision tracking telescope at the DESY test beam facilities. The high telescope track resolution enables precise studies of tracking efficiency, charge sharing and collection even within single pixel cells of the device under test.

This publication focuses on the improved performance and capabilities of the new pixel readout chip and summarizes results from test beam campaigns with both unirradiated and irradiated devices. The functionality of the chip design with its improved charge threshold, redesigned data transmission and buffering scheme has been verified.}


\begin{document}
\section{Introduction}\label{sec:intro}

The CMS pixel detector is the innermost component of the tracking system of the CMS experiment~\cite{cms}. Its main purpose is the precise measurement of the first space points of charged particle trajectories, close to the \emph{pp} interaction point. Its performance and efficiency is crucial for precise secondary vertex resolution and thus e.g. b-jet tagging algorithms.

The present pixel detector was designed to operate under conditions of $40\,\mathrm{MHz}$ bunch crossing frequency and an instantaneous luminosity of $1\times10^{34}\,\mathrm{cm^{-2}s^{-1}}$. It provides coverage with three space points up to pseudorapidities of $\eta < |2.5|$ using three barrel layers (referred to as BPIX) and two forward disks per side (FPIX). The charge deposited in the silicon sensors is read out, discriminated, and stored by the front end electronics readout chip (ROC).

At twice the design design luminosity of the LHC of $2\times10^{34}\,\mathrm{cm^{-2}s^{-1}}$ the detector would be subject to severe inefficiencies. Therefore an upgrade of the CMS pixel detector is planned~\cite{tdr}, including a new ROC in order to maintain the efficiency of the current pixel tracker at the increased requirements imposed by high luminosities and pile-up.

\section{The Current and Upgraded Readout Chips}\label{sec:upgrade}

\begin{figure}[tbp]
  \centering
  \hspace*{\fill}%
  \subfigure[Layout of the ROC]{\label{fig:periphery:a}\includegraphics[width=.3\textwidth]{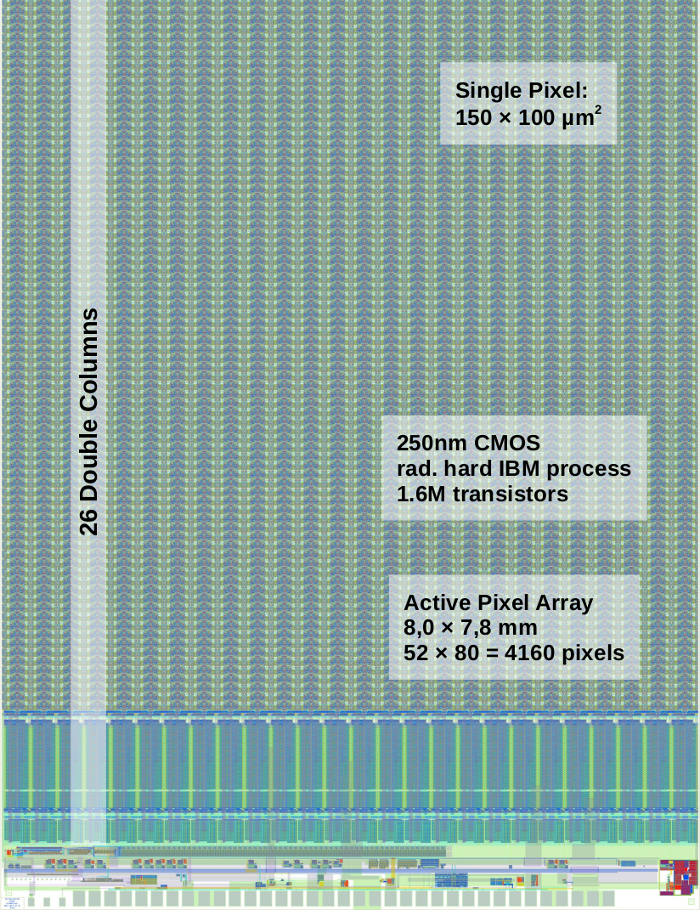}}%
  \hfill%
  \subfigure[Periphery footprint of current (left) and upgraded ROC (right)]{\label{fig:periphery:b}\includegraphics[width=.6\textwidth]{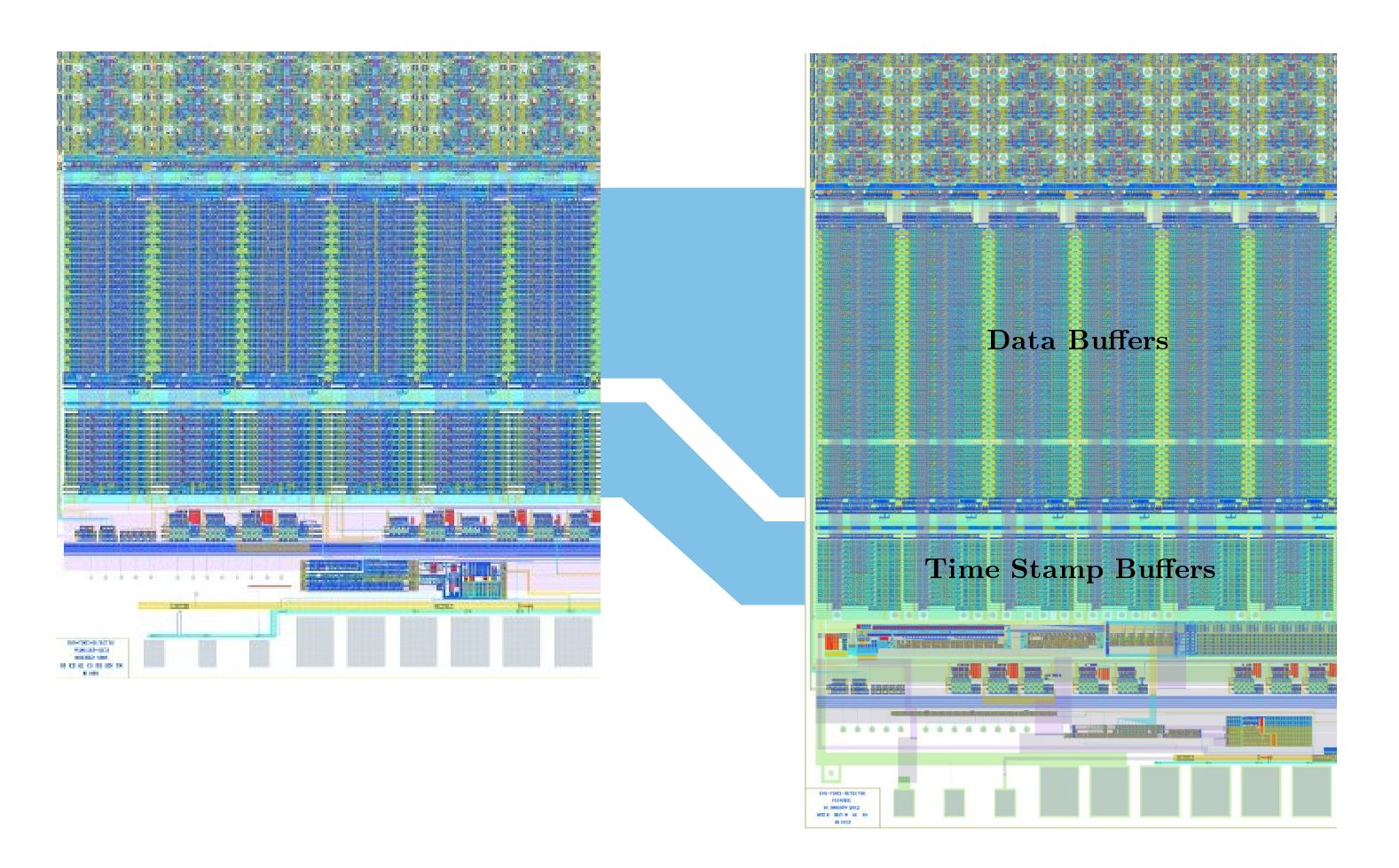}}%
  \hspace*{\fill}
  \caption{Schematics of the Phase-I Upgrade ROC (a) and comparison of the periphery footprint of the two ROC versions (b). The number of data buffers was increased from 32 to 80, the number of time stamp cells from 12 to 24.}
  \label{fig:periphery}
\end{figure}

The current pixel detector ROC is a radiation hard $250\,\mathrm{nm}$ process silicon chip which is bump-bonded to a \emph{n$^+$-in-n} silicon sensor. It comprises $4160$ pixels, each with its own shaper and comparator circuits and individually adjustable per-pixel charge thresholds for zero-suppression. The pixel rows and columns have a pitch of $100\,\mathrm{\upmu m}$ and $150\,\mathrm{\upmu m}$ respectively, and are grouped in 26 double columns with 160 pixels each as shown in Figure~\ref{fig:periphery:a}. Every double column is read out separately and has its own buffers. Pixels passing the threshold cut are read out, a time stamp is created, and the charge is stored in the double column periphery buffer cells.

While this chip was designed for and is performing well at instantaneous luminosities of $1\times10^{34}\,\mathrm{cm^{-2}s^{-1}}$, it would be subject to severe inefficiencies at higher values. The main causes for these ineffiencies would be time stamp and data buffer overflows, speed limitations in signal transmission from pixel cells to the periphery, and detector dead time during readout~\cite{vertex-beat}. Also the bandwidth of the analog $40\,\mathrm{MHz}$ level-encoded data links would not suffice to transfer all pixel hits in time. To overcome these limitations, a new ROC with significant improvements in various areas has been designed.

The Phase-I Upgrade ROC design is an evolution of the current ROC and the chip is fabricated in the same $250\,\mathrm{nm}$ process. The number of data buffers has been increased from 32 to 80 and the number of time stamp buffers from 12 to 24 cells, while reducing the size of the cells. This only slightly increases the periphery footprint and overall size of the chip, as shown in Figure~\ref{fig:periphery:b}, which allows to stay compatible with the current detector module design and to use the same sensor layout~\cite{sensor}. To further reduce dead time during the readout phase, an additional global readout buffer has been introduced, to which all pixel hits are transferred to from the double column buffers, prior to sending them over the data links.

The analog data link was replaced with a fully digital $160\,\mathrm{MHz}$ data link which roughly doubles the available bandwidth for data transmission. For this an additional $8\,\mathrm{bit}$ analog-to-digital converter (ADC) is necessary to encode the charge information for every pixel hit.

The individual pixel cells underwent further development which comprises a faster comparator allowing for lower charge thresholds and reduced time-walk effects as well as a reduction of the internal cross-talk. Both changes aim towards better position resolution as well as increased longevity of the detector due to a larger window for operational parameters.

While the current ROC would suffer from $16\%$ data loss at the innermost layer at instantaneous luminosities of $2\times10^{34}\,\mathrm{cm^{-2}s^{-1}}$ for $25\,\mathrm{ns}$ bunch spacing, simulations show that the design changes and improvements planned for the Phase-I Upgrade mitigate these effects resulting in an expected layer 1 data loss below $2.4\%$ for $25\,\mathrm{ns}$ LHC bunch spacing~\cite{vertex}.

\section{The DESY Testbeam Setup and Beam Telescope}\label{sec:tel}

Many test beam measurements with the evolved ROC design were carried out in order to verify the design, tune operation parameter settings, and ensure good performance of the new detector.

The test beam campaigns were carried out at the DESY test beam facility using positrons from the DESY synchrotron. The actual test beam is produced via conversion. The accelerator beam crosses a carbon fiber producing bremsstrahlung, which is guided onto copper targets where it undergoes pair production. Using a magnet, either electrons or positrons can be selected in an energy range of $1-6\,\mathrm{GeV}$. The resulting test beam has a divergence of about $1\,\mathrm{mrad}$, an energy spread of $\sim5\%$, and instantaneous particle rates of a few $\mathrm{kHz/cm^2}$ can be reached.

\begin{figure}[tbp]
  \centering
  \includegraphics[width=.9\textwidth]{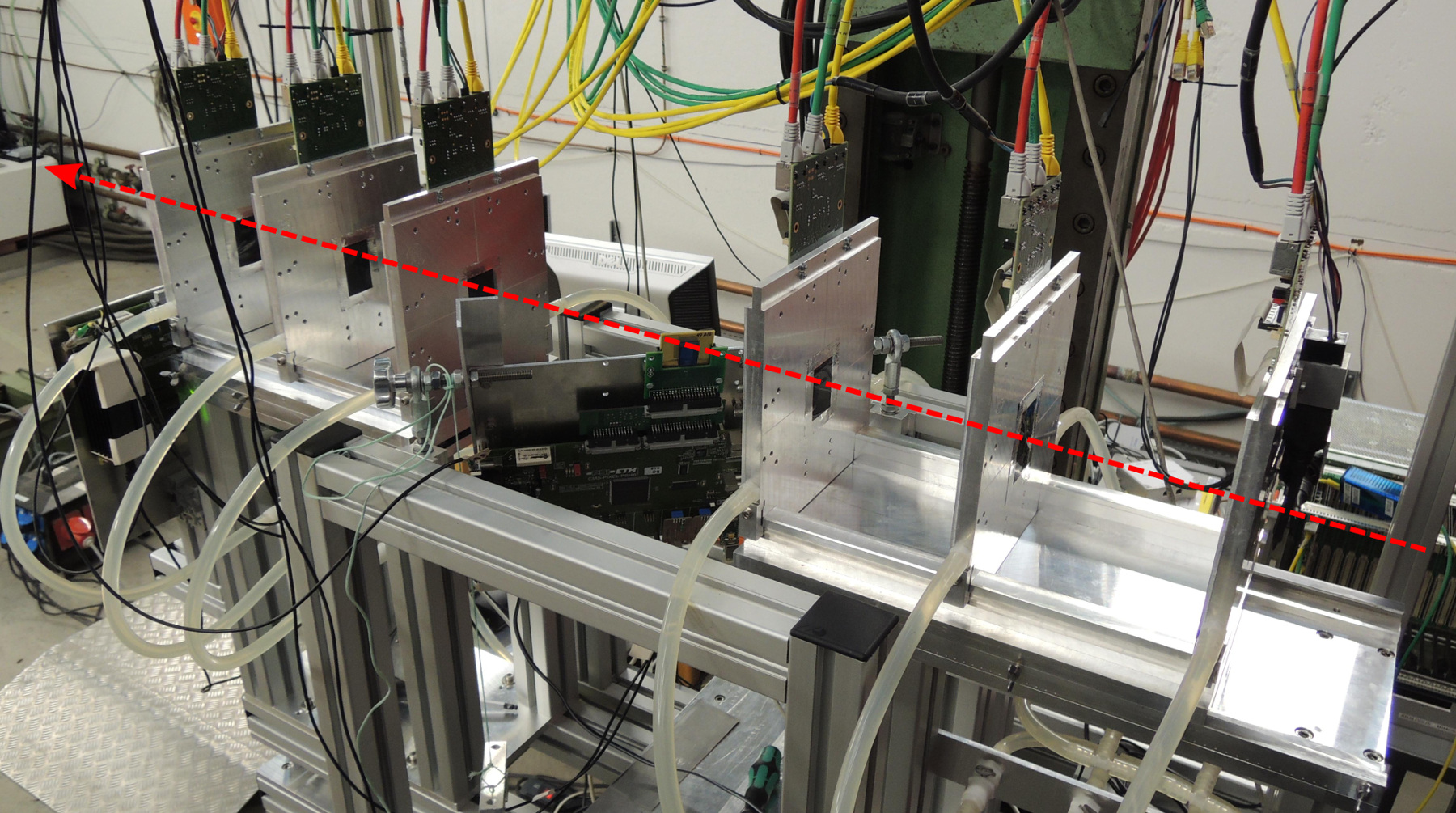}
  \caption{The DATURA beam telescope with a CMS pixel device mounted as DUT. The picture shows the six Mimosa26 planes with readout and cooling attached, the rotatable hinge with CMS ROC and readout electronics, and the two front scintillators. The test beam trajectory is marked with a red arrow.}
\label{fig:telescope}
\end{figure}

The ROCs operated as device under test (DUT) were placed between the two arms of an EUDET/AIDA-family pixel beam telescope~\cite{eudet} as shown in Figure~\ref{fig:telescope}. The beam telescope features six planes of Mimosa26 MAPS devices~\cite{mimosa} covering an area of $2\times1\,\mathrm{cm^2}$. These devices have quadratic pixels with a pitch of $18.4\,\mathrm{\upmu m}$ and are thinned to $50\,\mathrm{\upmu m}$ to reduce the effect of multiple scattering and thus improve the track resolution. The DUT was attached to a hinge allowing rotation around two axes in the beam. This enables studies of the detector behaviour with inclined particle tracks producing multi-pixel clusters.

In addition to the DUT in the center of the six Mimosa26 planes, a second CMS pixel ROC was placed downstream at the end of the beam telescope as a timing reference. This is necessary for efficiency measurements due to the rolling shutter readout of the Mimosa26 devices with an integration time of $\mathcal{O}(100\,\mathrm{\upmu s})$ and the resulting track multiplicity in the beam telescope.

The experimental setup was triggered using the signal coincidence of four scintillators placed up- and downstream of the beam telescope.

With the experimental setup descibed above a track extrapolation error of about $4.8\,\mathrm{\upmu m}$ at the position of the DUT can be achieved. This allows not only for determining the tracking efficiency of the DUT but also detailed studies of charge collection behaviour in single pixel cells (Section~\ref{sec:cce}).

\section{Readout Chip Characterization}

Using the experimental setup descibed in the previous section, several ROCs have been tested for their tracking efficiency, position resolution, and threshold characteristics. In addition, irradiated devices have been measured in order to characterize the performance after receiving their expected lifetime dose.

\subsection{Tracking Efficiency, Position Resolution}

The tracking efficiency of the Phase-I Upgrade ROC is measured to be $99.6\%$ by first selecting beam telescope tracks within the correct trigger window using the timing reference detector. The tracks are then matched to clusters on the DUT. The efficiency depends on the number of tracks in the beam telescope and the resulting ambiguities. Additionally requiring only a single telescope track in every event, a tracking efficiency of $99.9\%$ can be achieved.

\begin{figure}[tbp]
  \centering
  \hspace*{\fill}%
  \subfigure[Resolution as function of track incidence angle]{\label{fig:resolution:a}\includegraphics[width=.45\textwidth]{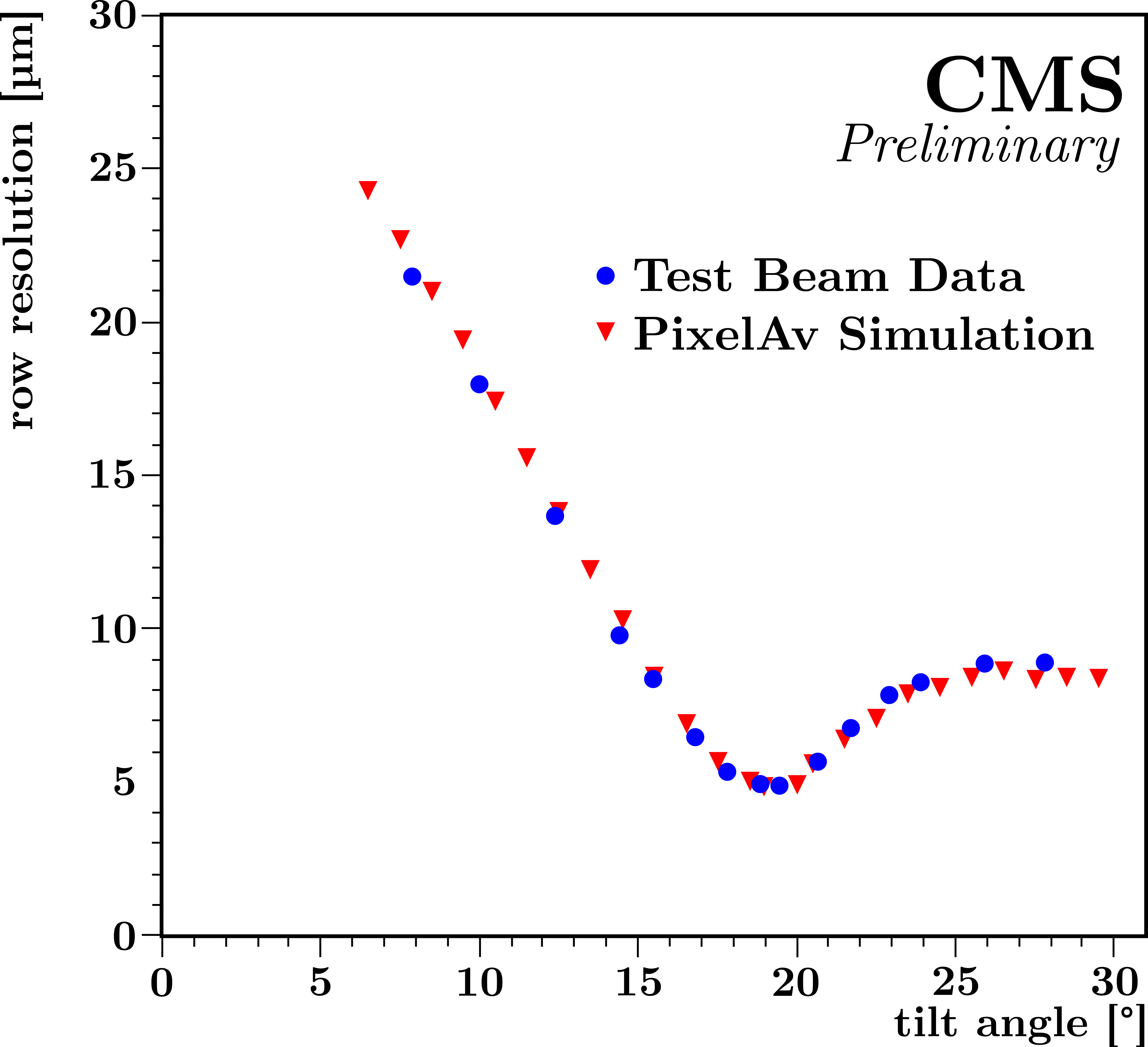}}%
  \hfill%
  \subfigure[Resolution as function of charge threshold]{\label{fig:resolution:b}\includegraphics[width=.45\textwidth]{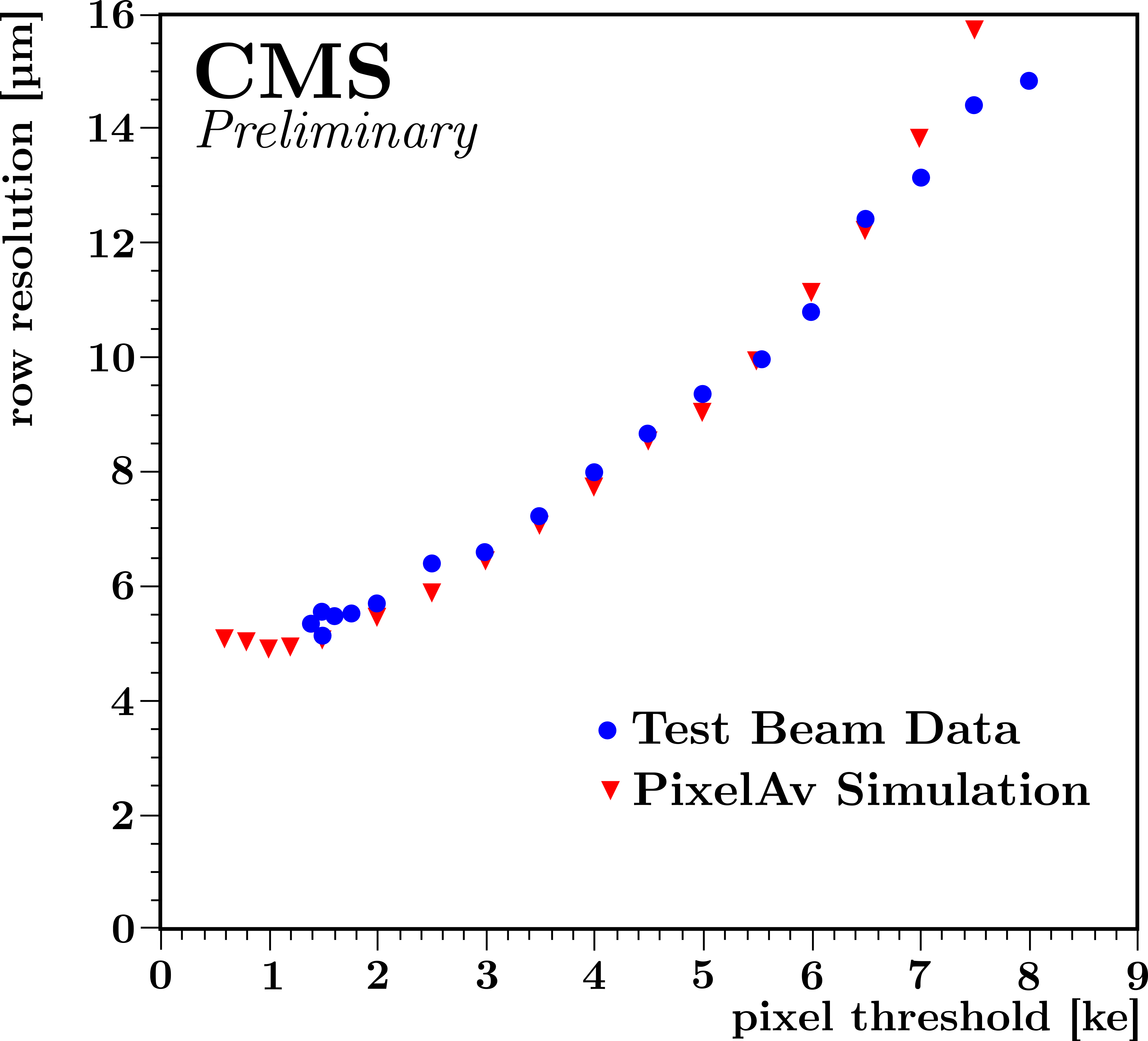}}%
  \hspace*{\fill}
\caption{Position resolution as function of (a) the track incident angle (mapping to the Lorentz angle in CMS) and (b) the per-pixel charge threshold in electrons. The plots show both test beam data (circles) and simulation (triangles).}
\label{fig:resolution}
\end{figure}

Using the rotatable hinge one can measure the position resolution at different particle track incidence angles. Tilting the device around the pixel row axis simulates the Lorentz angle by which charge carriers are deflected inside the sensor by the $3.8\,\mathrm{T}$ magnetic field of the CMS solenoid. The pixel size is carefully chosen such that this results in optimal charge sharing behaviour between two pixel cells and thus yields the best possible position resolution. Figure~\ref{fig:resolution:a} shows the position resolution across pixel rows with $100\,\mathrm{\upmu m}$ pitch achieved at different tilt angles. The best position resolution of about $5\,\mathrm{\upmu m}$ is reached at the optimal tilt angle of $19.5^{\circ}$ which is determined by the pixel geometry\footnote{atan(pixel width/sensor thickness) = atan($100\,\mathrm{\upmu m}/285\,\mathrm{\upmu m}) \approx 19.5^{\circ}$}. The telescope track extrapolation error has been subtracted. This measurement can be used to study the impact of charge carrier trapping on the position resolution. With higher bulk damage in the silicon sensor the Lorentz angle decreases and thus the position resolution deteriorates.

The position resolution also strongly depends on the charge threshold. Edge pixels of clusters receiving only a small portion of the overall deposited charge are more likely to be cut off with a higher threshold. Figure~\ref{fig:resolution:b} shows the dependence of the row resolution on the applied pixel threshold in units of $1000$ electrons. The current pixel ROCs are operated at a minimal threshold of about $3500\,\mathrm{e}$ where a resolution of $7\,\mathrm{\upmu m}$ can be achieved. Lowering the threshold to $1500\,\mathrm{e}$ results in a row resolution of $5\,\mathrm{\upmu m}$, and also increases the longevity of the detector. With increasing radiation damage, less charge will be collected at the implants which can partly be compensated by applying a lower pixel charge threshold.

The test beam data was verified by comparing the results to detailed simulation studies performed using the CMS PixelAv package~\cite{pixelav}. The package includes effects of scattering in silicon, Delta rays as well as charge carrier drift and diffusion. Timing and threshold effects are simulated to model the ROC response. The resolution as function of track incidence as well as charge threshold settings is in good agreement with the test beam data as can be seen from Figure~\ref{fig:resolution}.

\subsection{Charge Collection Efficiency}\label{sec:cce}

The charge collection efficiency as function of the track position within a single pixel cell of the CMS pixel ROC can be measured using the high telescope track resolution of $4.8\,\mathrm{\upmu m}$ at the DUT (compare Section~\ref{sec:tel}).

\begin{figure}[tbp]
  \centering
  \hspace*{\fill}%
  \subfigure[Sensor photograph of four BPIX pixel cells]{\label{fig:puc:a}\includegraphics[width=.45\textwidth]{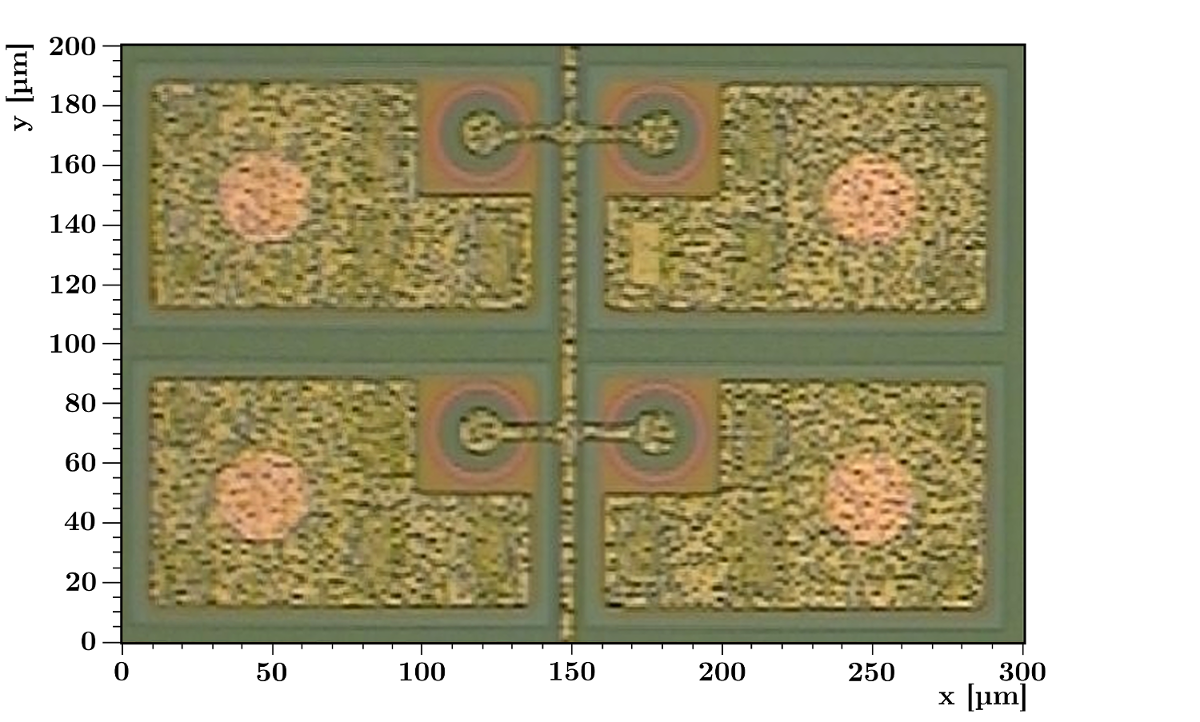}}\hfill%
  \subfigure[Perpendicular track incidence: $\theta = \phi = 0$]{\label{fig:puc:b}\includegraphics[width=.45\textwidth]{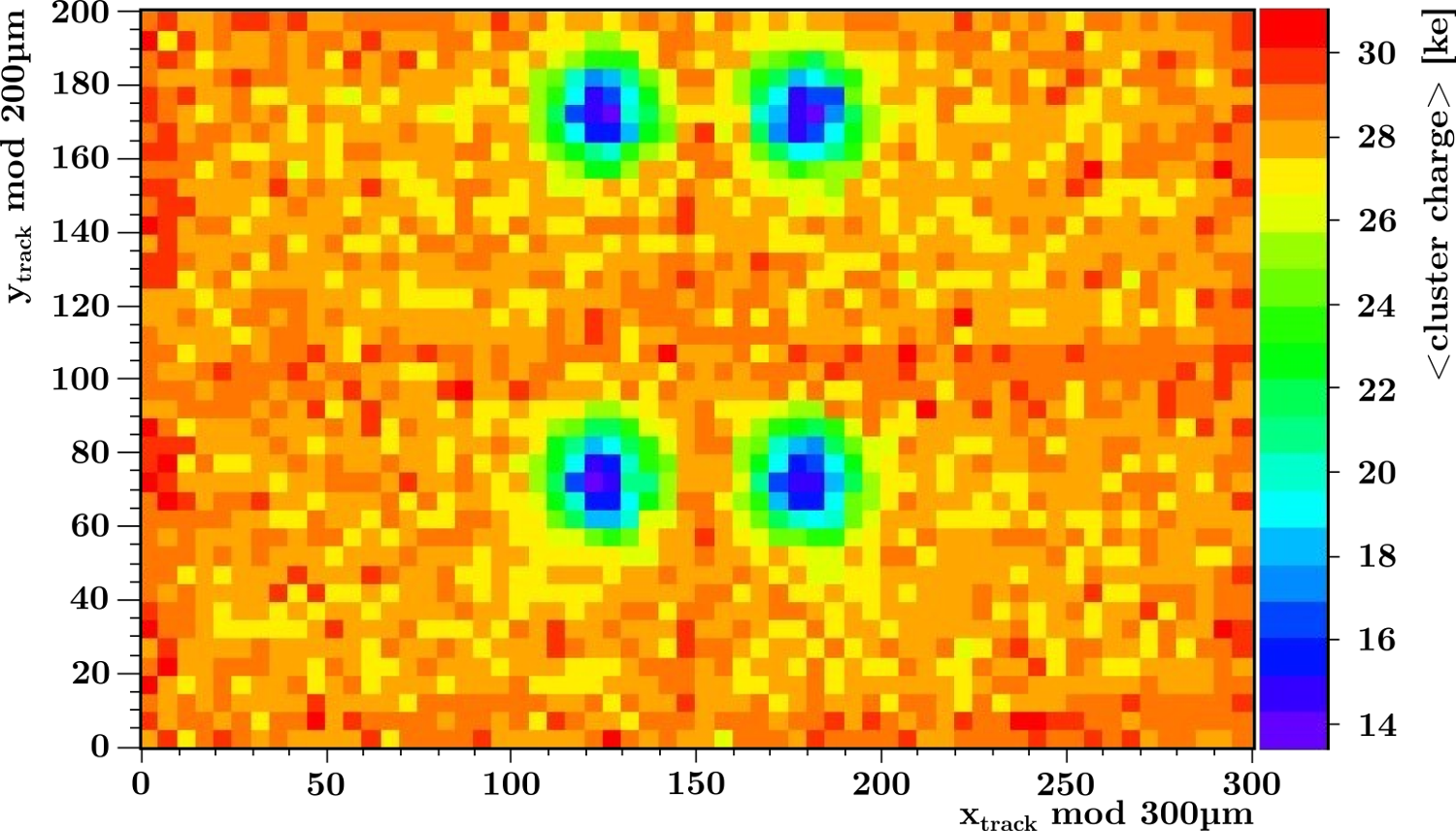}}%
  \hspace*{\fill}

  \hspace*{\fill}%
  \subfigure[Rotated around row axis: $\theta = 0^{\circ} \quad \phi = 19^{\circ}$]{\label{fig:puc:c}\includegraphics[width=.45\textwidth]{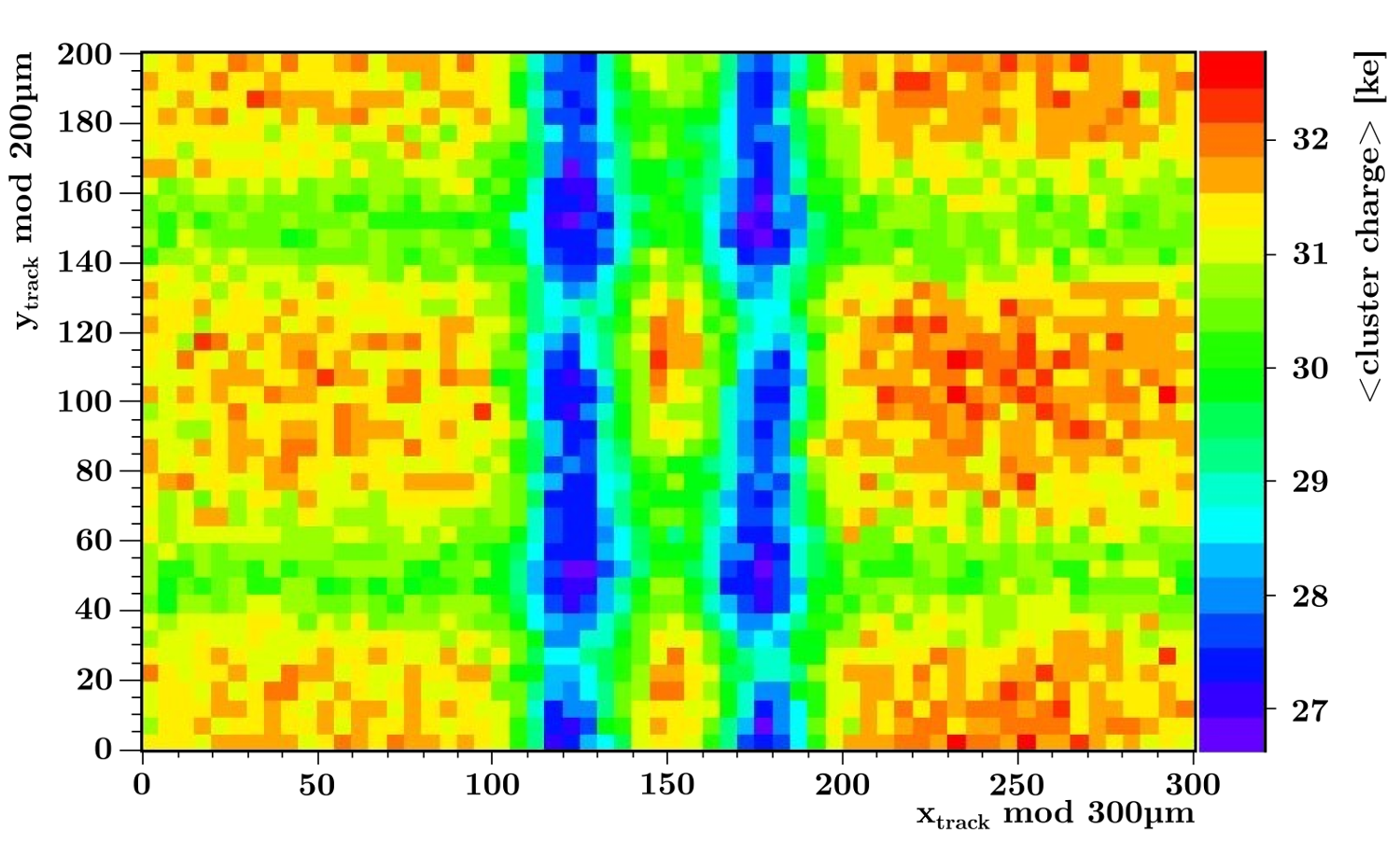}}%
  \hfill%
  \subfigure[Rotated around two axes: $\theta = 28^{\circ} \quad \phi = 19^{\circ}$]{\label{fig:puc:d}\includegraphics[width=.45\textwidth]{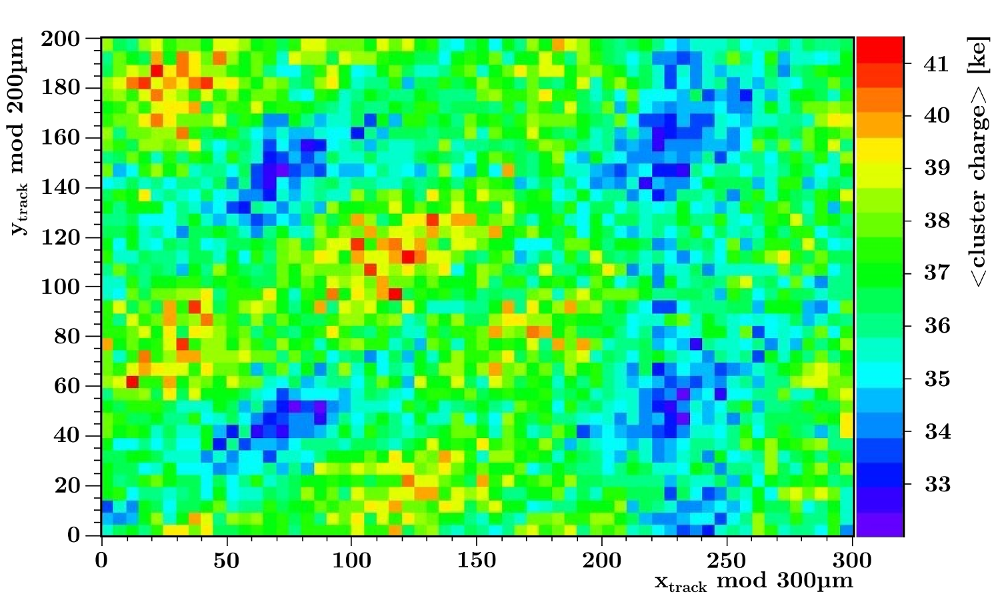}}%
  \hspace*{\fill}
\caption{Charge maps showing the total cluster charge as function of the track position within pixel cells. To increase statistics, tracks from the whole sensor area have been folded into the four pixel cells (absolute track position modulo $\mathrm{pitch}\times2$). The \emph{z}-axes of plots (b)--(d) have different scales to emphasize the differences.}
\label{fig:puc}
\end{figure}

Figure~\ref{fig:puc:a} shows a photograph of four BPIX pixel cells in the same double column. Visible are the implants as well as the bump pad and the punch-through grid. The latter is set to ground potential to prevent pixels with bad bumps to float to bias voltage. This alters the electric field in the sensor bulk below and leads to reduced charge collection as can be seen in Figure~\ref{fig:puc:b}. With particles incident perpendicular to the sensor surface only half of the charge deposited in the pixel is collected below the punch-through dots.

However, with the magnetic field inside the CMS detector Lorentz drift has to be taken into account. This smears out the effect of the punch-through dots as shown in Figure~\ref{fig:puc:c} with the charge loss being reduced to about $15\%$. Figure~\ref{fig:puc:d} also takes into account rotations around the second axis (which represents different pseudorapitidies along the $z$-axis of CMS) which further mitigates the effect, and no impact on the tracking efficiency was observed.

\subsection{Performance of Irradiated Devices}

To verify the detector performance after several years of operation, CMS pixel devices consisting of a Phase-I Upgrade ROC and a bump-bonded sensor were irradiated up to the expected lifetime dose of layer 4 after $500\,\mathrm{fb^{-1}}$ integrated luminosity. The devices were irradiated using the $24\,\mathrm{GeV}$ proton beam at the CERN Proton Synchrotron (PS), the maximum dose received is $13\,\mathrm{Mrad}$ which corresponds to a fluence of $\phi \approx 2.3\times10^{14}\,\mathrm{n_{eq}/cm^2}$. The devices were characterized in the DESY test beam both before and after irradiation and operated at a charge threshold of $1500\,\mathrm{e}$.

\begin{figure}[tbp]
  \centering
  \hspace*{\fill}%
  \subfigure[\emph{x-y} map of the DUT tracking efficiency]{\label{fig:irrad:a}\includegraphics[width=.5\textwidth]{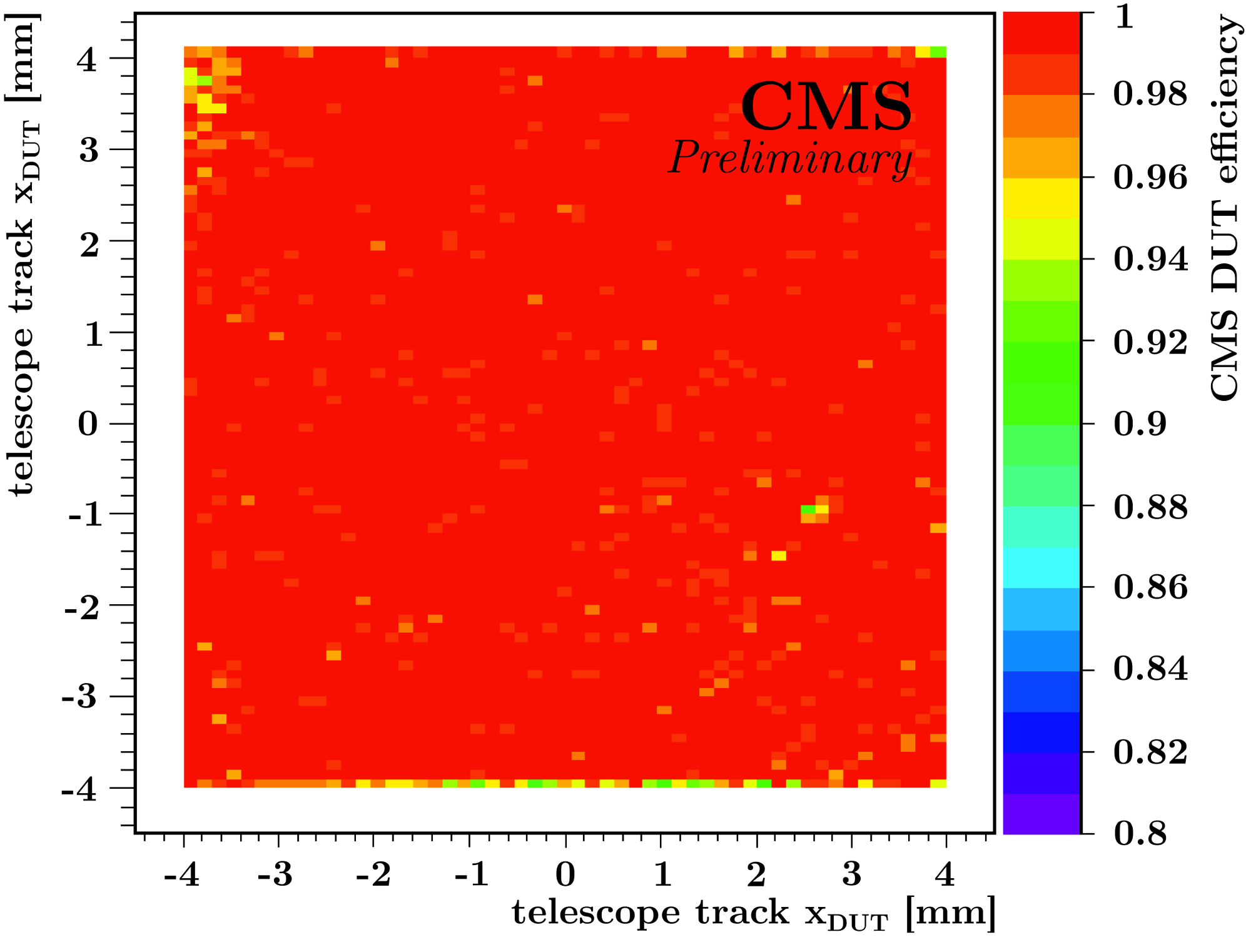}}%
  \hfill%
  \subfigure[Track residual along rows on the DUT]{\label{fig:irrad:b}\includegraphics[width=.38\textwidth]{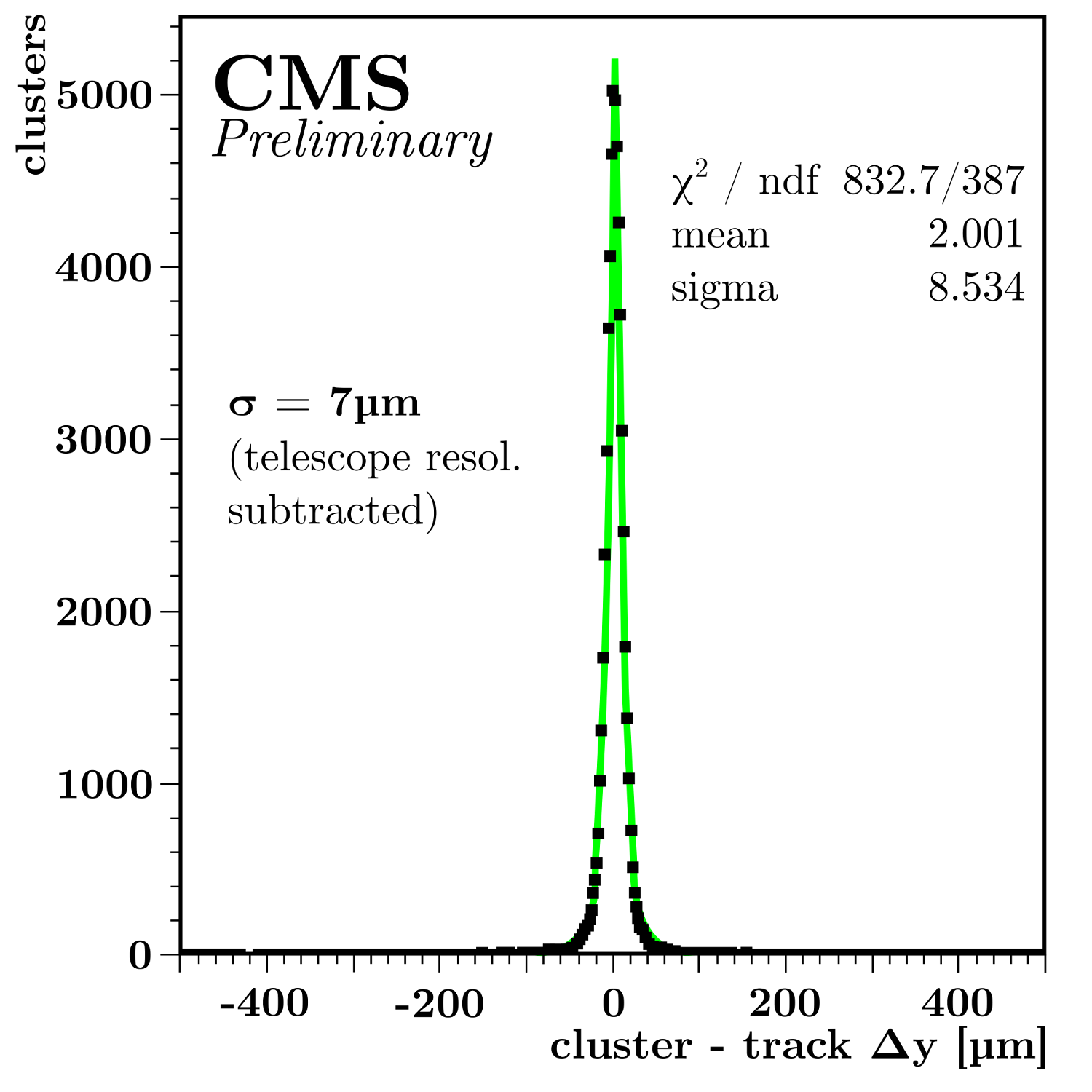}}%
  \hspace*{\fill}
\caption{Performance figures for the Phase-I Upgrade ROC after irradiation with a dose of $13\,\mathrm{Mrad}$: (a)~tracking efficiency, and (b)~position resolution along the pixel rows with $100\,\mathrm{\upmu m}$ pitch.}
\label{fig:irrad}
\end{figure}

Figure~\ref{fig:irrad:a} shows an \emph{x-y} map of the tracking efficiency of the irradiated device in the DESY test beam, measured with the same experimental setup as described before. Even after irradiation the overall efficiency is above 98\% and no significant pattern across the chip can be seen. Only in some small regions the tracking efficiency degraded to about 90\% which might be due to inhomogenities during the irradiation process or not optimally adjusted pixel thresholds after irradiation. The position resolution is still very good with $7\,\mathrm{\upmu m}$ in row direction as shown in Figure~\ref{fig:irrad:b}.

\begin{figure}[tbp]
  \centering
  \includegraphics[width=.45\textwidth]{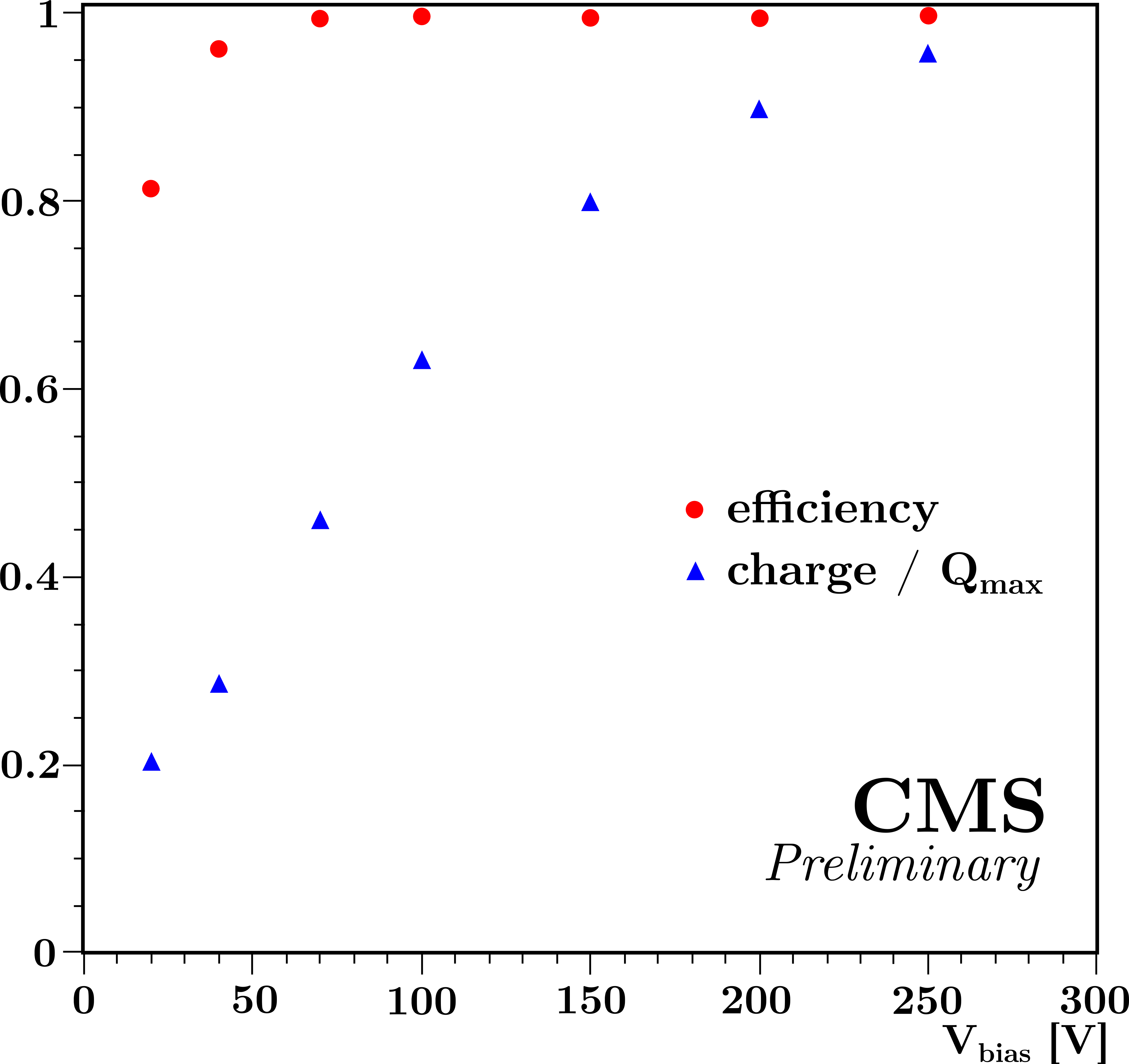}
  \caption{Charge collection efficiency, normalized to the maximum cluster charge (triangles) and tracking efficiency (circles) of the irradiated Phase-I Upgrade ROC as function of applied bias voltage. Even with an underdepleted sensor, the chip operated above 98\% tracking efficiency while collecting only half of the charge deposited.}
  \label{fig:exit}
\end{figure}

With increasing radiation damage the sensor will only be partially depleted due to limits on the power dissipation and the maximum bias voltage applied. Due to the \emph{n}-type bulk material, the sensor undergoes type inversion during irradiation and thus the depleted region is located on the implant side which allows to collect the charge even in partially depleted sensors. To emulate this effect, the bias voltage applied to the DUT was varied between full depletion at $250\,\mathrm{V}$ and $20\,\mathrm{V}$ to deplete volumes of different thicknesses and thus varying the amount of charge collected. Figure~\ref{fig:exit} shows both the cluster charge normalized to the maximum charge collected and the tracking efficiency of the device as function of the bias voltage.

The collected cluster charge drops quickly for bias voltages below $150\,\mathrm{V}$. At around $80\,\mathrm{V}$ only half of the charge is collected, and a bias voltage of $20\,\mathrm{V}$ yields about 20\% of the full cluster charge. The efficiency remains above 98\% down to a bias voltage of around $70\,\mathrm{V}$ and is still at about 80\% with the low charge collection efficiency of 20\% at $20\,\mathrm{V}$ bias voltage. This behaviour significantly extends the lifetime of the detector.

\section{Conclusion}\label{sec:conclusion}

An upgrade of the CMS pixel detector as proposed in~\cite{tdr} will ensure the good performance of the detector at the increased requirements imposed by higher LHC luminosity of $2\times10^{34}\,\mathrm{cm^{-2}s^{-1}}$ which is likely to be reached before 2018. The new front end readout chip for the pixel detector implements many features mitigating the effect of higher pixel occupancies.

Additional data and time stamp buffer cells, a more efficient buffering scheme, and a fully-digital $160\,\mathrm{MHz}$ data link allow for more pixel hits to be recorded, stored, and transmitted, while redesigned components in the pixel cells like a faster comparator provide a larger working range by lowering the charge threshold and thus increase the longevity of the detector.

Test beam measurements at the DESY synchrotron verified the design and demonstrate the improved performance of the chip. The position resolution as well as tracking efficiency after irradiation profit from the lower charge threshold. The charge collection efficiency has been studied as function of the track impact point within single pixel cells.

The Phase-I Upgrade ROC has shown its capability to be operated efficiently in the environment expected at higher LHC luminosities.


\end{document}